# Quantum physics and the beam splitter mystery

François Hénault

Institut de Planétologie et d'Astrophysique de Grenoble

Université Joseph Fourier, Centre National de la Recherche Scientifique

B.P. 53, 38041 Grenoble – France

## ABSTRACT

Optical lossless beam splitters are frequently encountered in fundamental physics experiments regarding the nature of light, including "which-way" determination of light particles, N. Bohr's complementarity principle, or the EPR paradox and all their measurement apparatus. Although they look as common optical components at first glance, their behaviour remains somewhat mysterious since they apparently exhibit stand-alone particle-like features, and then wave-like characteristics when inserted into a Mach-Zehnder interferometer. In this communication are examined and discussed some basic properties of these beamssplitters, both from a classical optics and quantum physics point of view. Herein some convergences and contradictions are highlighted, and the results of a few emblematic experiments demonstrating photon existence are discussed. An alternative empirical model in wave optics is also proposed in order to shed light on some remaining questions.

**Keywords:** beams-splitter, interferometer, Mach-Zehnder, photon, correlation, coincidence counting

## 1 INTRODUCTION

Quantum physics theory, or more precisely quantum optics as is dealt with in the present paper, is nowadays illustrated by a rich set of exemplary experiments demonstrating the behaviour of photons as particles and the existence of non-intuitive physical properties such as paths undistinguishability or non-locality of those particles. One may cite for example:

- An illustrative experiment described by Grangier, Roger and Aspect (GRA), being built around a single beam splitter (BS) revealing corpuscle behaviour through analysis of the transmitted and reflected beams coincidence counts, and later integrated into a Mach-Zehnder (MZ) interferometer and still producing interference fringes, i.e. wave-behaving evidence [1].

- The Hong-Ou-Mandel (HOM) effect demonstrating that two photons simultaneously impinging on both opposite sides of a BS cannot cross or be reflected simultaneously [2].

- The question of escaping the Einstein, Podolsky and Rosen (EPR) paradox, who has been the subject of a long suite of successful tests, where polarization correlations or anti-correlations are evidenced by use of rotating beamsplitters [3-4].

It must be emphasized that all these experiments, as well as their numerous variants generated throughout the years, involve the use of one or several BS, which appear consequently as crucial objects supporting our current understanding of the nature of light. Thus it is of some interest to review the characteristics of these optical components being often considered as common at first glance, but actually exhibit either particle-like or wave-like features. There is the starting point to the present communication, which is organized as follows:

- Section 2 provides a quick overview of the BS theory, both from the quantum physics point-of-view where it is essentially handled as a macroscopic operator (§ 2.1), and then following cclassical wave optics in a more tangible way of understanding (§ 2.2).

- Section 3 is especially focused at basic quantum optics experimental setups aiming at demonstrating the existence of the photon and dual nature of light, namely the Hanbury Brown and Twiss experiment used in coincidence counting mode (§ 3.1) and the Mach-Zehnder interferometer (§ 3.2). For each case the experimental results are discussed, and compatibility with classical wave theory is discussed.

# 2 BEAMSPLITTER THEORETICAL MODELS

In this section are reviewed the two major physical interpretations of light interacting with a beamsplitter, successively following quantum optics formalism (§ 2.1) and classical wave optics involving Fabry-Perot interference effect (§ 2.2). It is concluded that both point-of-views are essentially convergent (§ 2.3).

## 2.1 Quantum physics

Common use in quantum physics is to consider a macroscopic, "black-box beamsplitter" having two input ports n° 1 and 2 and two output ports n° 1' and 2', linked by a matrix relationship and satisfying to certain physical or mathematical necessary conditions. This BS matrix $M_{BS}$ is usually built from the physical quantities depicted in Figure 1, writing as:

$$\begin{bmatrix} A'_1 \\ A'_2 \end{bmatrix} = M_{BS} \begin{bmatrix} A_1 \\ A_2 \end{bmatrix} = \begin{bmatrix} t_{11} & r_{21} \\ r_{12} & t_{22} \end{bmatrix} \begin{bmatrix} A_1 \\ A_2 \end{bmatrix}, \tag{1}$$

where $A_1$ and $A_2$ are the input amplitudes in ports 1 and 2, $A_1$' and $A_2$' output amplitudes in ports 1' and 2', and the complex coefficients $t_{ij}$ and $r_{ij}$ are the amplitude transmission and reflexion factors from and towards different BS ports. One may note that the representation of output amplitudes $A_1$' and $A_2$' in Figure 1, respectively splitted into $A_{T1}$ and $A_{R2}$ and $A_{T2}$ and $A_{R1}$, may look somewhat ambiguous, eventually challenging P. A. M. Dirac's famous statement about indivisible photons. However the contradiction is only apparent; because quantum theory assumes that each photon can only go one way, i.e. it is either transmitted or reflected, but not both simultaneously. Moreover this representation is also valid for the wave optics model described in sub-section 2.2.

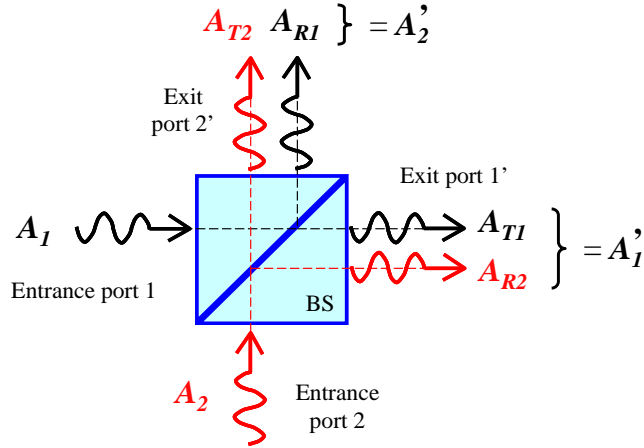

Figure 1: Quantum physics scheme of the black-box beamsplitter.

The present paper is essentially focused at lossless beamsplitters, meaning that the incident light beams do not suffer from photon loss either at BS crossing or reflecting. According to quantum optics theory the lossless property translates into the conditions:

$$\begin{aligned} A'_1 A'^*_1 + A'_2 A'^*_2 &= A_1 A^*_1 + A_2 A^*_2 = 1 \quad and: \\ A'_1 A'^*_2 + A'_2 A'^*_1 &= 0, \end{aligned} \tag{2}$$

where * denotes complex conjugates. After elementary manipulation of the complex coefficients $t_{ij}$ and $r_{ij}$, the BS matrix in relation 1 is found to be:

$$M_{BS} = \frac{\exp(i\varphi_0)}{\sqrt{2}} \begin{bmatrix} i & 1 \\ 1 & i \end{bmatrix}, \qquad (3a)$$

with $i = \sqrt{-1}$ and $\varphi_0$ is an arbitrary phase constant usually set to zero. Hence both transmitted light beams undergo a phase-shift equal to π/2, while the reflected beams are unaffected. There are several different ways to ascertain the existence of this phase-shift and demonstrating Eq. 3a, involving more or less abstract notions of quantum physics. Here are only cited a few of them:

- Degiorgio used the principle of energy conservation into a Michelson interferometer to justify a π/2 phase difference between both BS output ports [5], later imitated by Ou and Mandel [6].

- Zeilinger introduced the matrix formalism and generalized Degiorgio's demonstration by stating that the lossless BS matrix must be unitary [7]. The same approach was developed and clarified by Holbrow, Galvez and Parks [8], mentioning that the unitarity condition derives from conservation of probabilities.

- Fearn and Loudon presented a modal and field-quantized[1] theory of photo-count probabilities generated by two photons impinging in input ports n° 1 and 2, however their BS model is mostly similar to the previous ones [9].

- Campos, Saleh and Teich pointed out that "energy conservation alone is not sufficient to determine a standard beam-splitter transformation." Their complex amplitudes $A_1$, $A_2$, $A_1'$ and $A_2'$ are interpreted as the bosons annihilation operators at the input and output ports of the BS, having to obey boson commutation rules. Consequently, $M_{BS}$ should belong to the SU(2) subgroup of unitary operators [10].

- Note finally that Campos *et al* as well as a few other authors (see e.g. Ref. [11]) are using a different form of BS matrix, where the *i* factor is omitted as in Eq. 3b. It seems however that this last representation should be restricted to the case of half-silvered mirrors, as will be discussed in sub-sections 2.2 and 3.2.1.

$$M_{BS} = \frac{1}{\sqrt{2}} \begin{bmatrix} 1 & 1 \\ -1 & 1 \end{bmatrix}. \qquad (3b)$$

This quick tour of BS quantum theory is far from being complete. Nevertheless and whatever may be the purest theoretical demonstration, it may be concluded that:

1) Quantum optics essentially provides black-box models of the beamsplitter. They all agree on the existence of a π/2 phase-shift, even if its sign and precise location (on the transmitted or reflected beams) are uncertain. Such inconsistencies, however, are not critical for what concerns the respect of energy conservation principle.

2) Most authors (but not all) seem to put emphasis on the "lossless" or non-absorbing property of the BS, relegating its symmetric design at a position of secondary importance. Then it is unclear if the latter is a necessary condition for achieving the π/2 phase-shift. The wave optics model described in the next sub-section should provide clearer answer to that question.

3) Similarly, it must be emphasized that the here above formalism is not well suited to easy introduction of non ideal beamsplitters having non-symmetric or absorbing properties, while this is possible by using the wave optics formalism.

## 2.2 Classical wave optics

The fact that the transmitted and reflected beams of a BS are phase-shifted by π/2 has sometimes been considered with a bit of skepticism by certain optical specialists (including the author some time ago). One possible reason is that there exists a wide variety of such optical components as sketched in Figure 2[2]. When visiting an optical workshop, one may find for example cube BS (leftmost panel in the figure), half-silvered mirrors (central panel) or pellicle BS made of a single, extremely thin plastic layer (right panel). From these three examples, only the first and last ones are in fact symmetric and easily tractable using a classical formalism without developing a detailed model of optical coatings. Therefore the rest of this study shall essentially be focused at symmetric beamsplitters.

---

[1] Quantization involves the introduction of photon-annihilation and photon-creation operators, two pillars of quantum optics.
[2] More generally, any air-glass interface actually is a beamsplitter.

Here below is described a very classical model in wave optics, based on the Fabry-Perot (FP) interference effects of a single plane and parallel glass plate surrounded by air, thus forming a symmetric, sandwich-like configuration. This model is schematically illustrated in Figure 3, showing the BS parameters at media interfaces (left panel) and the multiple interference scheme (right panel). Scientific and geometric notations employed throughout the remainder of the paper are summarized in Table 1.

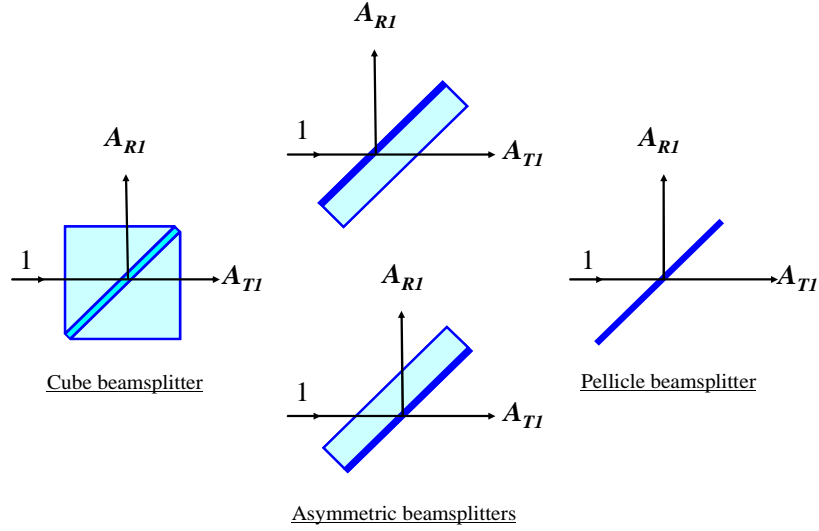

Figure 2: Different beamsplitter concepts. The input amplitude $A_1$ is normalized to 1 and output amplitudes are noted $A_{T1}$ and $A_{R1}$ in reference to Figure 1.

Table 1: Main employed scientific and geometric notations.

| | |
|---|---|
| $t_{12}$ | Real number, amplitude transmission factor from air to glass |
| $t_{21}$ | Real number, amplitude transmission factor from glass to air (always differing from $t_{12}$ according to Fresnel's relations) |
| $r_{12}$ | Real number, amplitude reflexion factor from air to air (external to glass) |
| $r_{21}$ | Real number, amplitude reflexion factor from glass to glass (internal to glass) |
| $\varphi$ | "Internal" phase-shift $\varphi = k\,\delta$ due to a single glass-crossing |
| $k$ | Wavenumber $k = 2\pi/\lambda$ of the electro-magnetic field, where $\lambda$ is its wavelength |
| $\delta$ | Optical path difference (OPD) $\delta = k\,n\,e\,\cos\theta$ due to a single glass-crossing |
| $n$ | Glass refractive index |
| $e$ | Glass plate thickness |
| $\theta$ | Internal incidence angle at glass-air interface |
| O and O' | Reference points for total OPD summation |

Assuming one single input amplitude $A_1 = 1$ in BS port n°1 and selecting O as the OPD reference point, one finds the following expressions of the transmitted and reflected amplitudes $A_{T1}$ and $A_{R1}$:

$$A_{T1} = \frac{t_{12}t_{21}\exp(i\varphi)}{1 - r_{21}^2 \exp(2i\varphi)} \quad (4a) \quad \text{and} \quad A_{R1} = \frac{r_{12} + r_{21}(t_{12}t_{21} - r_{12}r_{21})\exp(2i\varphi)}{1 - r_{21}^2 \exp(2i\varphi)}, \quad (4b)$$

where, following the previous notations: 
$$\varphi = k\,ne\cos\theta = \frac{2\pi}{\lambda}ne\cos\theta. \quad (5)$$

At this stage two important remarks should be made:

- For convenience, $A_{T1}$ and $A_{R1}$ are herein described as functions of the BS internal phase-shift $\varphi$. Practically, variations of $\varphi$ could either originate from the glass thickness $e$ or from the wavenumber $k$ (or wavelength $\lambda$) of the input beam. The last case is the most frequently encountered with FP interferometers, e.g. for spectroscopic applications.

- In classical textbooks, the expression of $A_{T1}$ (Eqs. 4 here above) generally omits the $\exp(i\phi)$ factor, implicitly choosing O' as OPD reference point for the transmitted amplitude, but still using O for the reflected beam. Taking this term into account is however fundamental for establishing the reality of the $\pi/2$ phase-shift, and may explain why it is not very well-known in classical optics.

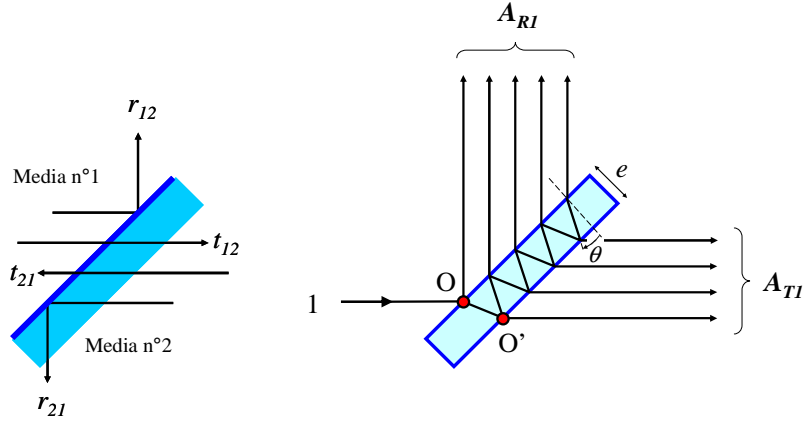

Figure 3: Fabry-Perot model of the symmetric beamsplitter. $t_{12}$, $t_{21}$, $r_{12}$ and $r_{21}$ are all real numbers.

The previous expression of $A_{R1}$ can be simplified when assuming non-absorbing (or lossless) media, and from classical properties of Fresnel's transitive and reflective coefficients, giving $r_{21} = -r_{12}$ and $t_{12}t_{21} - r_{12}r_{21} = 1$ [12-14]:

$$A_{R1} = -r_{21} \frac{1-\exp(2i\varphi)}{1-r_{21}^2 \exp(2i\varphi)} \tag{6}$$

The lossless BS condition also implies that $t_{12}^2 + r_{12}^2 = 1$ and $t_{21}^2 + r_{12}^2 = 1$. Then it is not very difficult to check that $|A_{T1}|^2 + |A_{R1}|^2 = 1$, thereby confirming the conservation of energy. The phase-shift between the reflected and transmitted beams $\phi_{R1} - \phi_{T1}$ is also obtained easily, determining the argument of the complex quantity $A_{R1}A_{T1}^*$. It is found that:

$$\phi_{R1} - \phi_{T1} = Arg[A_{R1}A_{T1}^*] = Arg[i\sin\varphi] = \pm\frac{\pi}{2}. \tag{7}$$

This result demonstrates that the phase-shift is always equal to one-fourth of the wave regardless of other optical phenomena such as wavelength-dependence or polarization of the electromagnetic field, therefore confirming the black-box approach followed by quantum mechanics. The analogy can be pushed further by rewriting the quantum BS matrix as follows:

$$M_{BS} = \frac{\exp(i\varphi)}{1-r_{21}^2\exp(2i\varphi)}\begin{bmatrix} t_{12}t_{21} & 2ir_{21}\sin\varphi \\ 2ir_{21}\sin\varphi & t_{12}t_{21} \end{bmatrix}. \tag{8}$$

The facts that the matrix multiplying factor is a complex number and the phase-shift $i$ is now attached to the reflected waves are not relevant. Of more interest is the necessary condition for achieving a perfect intensity balance between both BS output beams, writing:

$$\varphi = \pm \arcsin(t_{12}t_{21}/2r_{21}). \tag{9}$$

Practically, it can be realized at one single wavelength $\lambda$ and for a given glass plate thickness $e$: Then the BS intensity balance shall not be achromatic, independently of the relative phase-shift $\phi_{R1} - \phi_{T1}$. It must be noted that the previous results are in full agreement with FP interferometers and optical coatings theory [12].

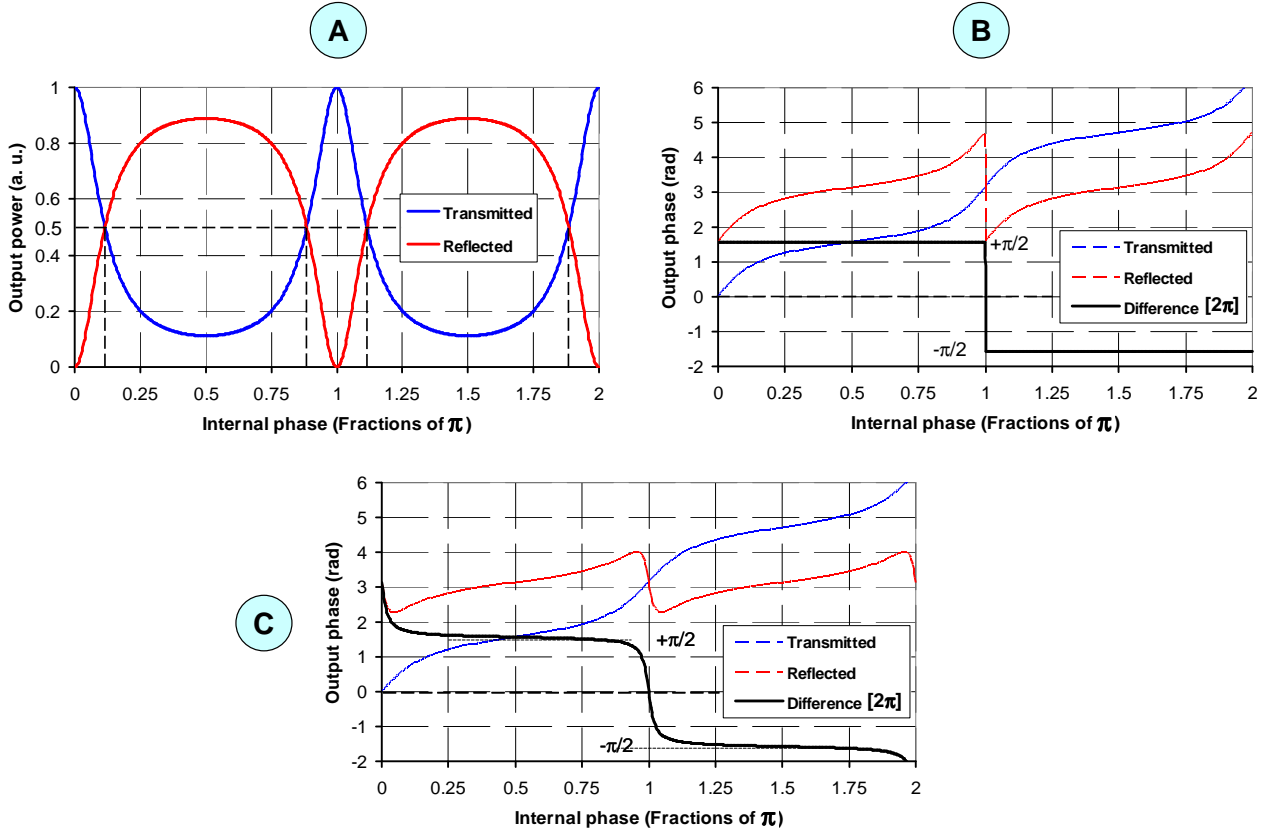

Figure 4: Transmitted and reflected intensities from a symmetric BS (panel A) and phases for the lossless and absorbing cases (panels B and C respectively).

Eqs. 4-5 also provide the basis for simplified numerical simulations of the BS properties, including the output intensities $I_{T1} = |A_{T1}|^2$, $I_{R1} = |A_{R1}|^2$ and their relative phase-shift $\phi_{R1} - \phi_{T1}$. This is illustrated in the panels A and B of Figure 4, where is considered the case of an intensity-balanced BS with $t_{12}t_{21} = 0.5$, $r_{21} = -r_{12} = \sqrt{2}/2$, and the internal phase $\varphi$ is varying between 0 to $2\pi$, resulting either from wavelength or glass plate thickness changes. The panel A actually reproduces typical spectral transmittance and reflectance curves of a FP interferometer (respectively shown in blue and red) complying with energy conservation rule $|A_{T1}|^2 + |A_{R1}|^2 = 1$ Equal BS output intensities $|A_{T1}|^2 = |A_{R1}|^2 = 0.5$ are achieved when $\varphi \approx 0.115\pi$ [$\pi$] and $\varphi \approx 0.885\pi$ [$\pi$] in agreement with Eq. 9. In panel B the relative phase difference is always found equal to $\pi/2$ or to $-\pi/2$ as a function of the internal phase $\varphi$ and can consequently be considered as achromatic. The panel C is illustrating the case of an absorbing BS whose transmitting and reflecting coefficients are such that $t_{12}t_{21} = 0.45$ and $-r_{12}r_{21} = 0.45$, corresponding to an absorption of 10%. Here the relative phase difference itself becomes a continuous function of the BS internal phase, being equal to $\pm\pi/2$ only when $\varphi = \pi/2$ or $\varphi = 3\pi/2$, both values at which the output intensities are unbalanced (i.e. $I_{R1} \neq I_{T1}$). It must be noted that this case is usually not covered by BS quantum models.

## 2.3 Conclusion

Having briefly reviewed the quantum and classical optical models of the beamsplitter, one may finally conclude that:

- It can be reasonably conjectured that the classical optical model can be generalized to all types of beamsplitters, including, pellicle BS, cube BS where the input/output and intermediate media have two different refractive indices $n_1$ and $n_2$ (with $n_1 < n_2$ or $n_1 > n_2$), or multi-layer semi-reflective coatings. Those beamsplitters may or may not be symmetric or lossless, thus rendering the classical model much more powerful than the quantum one.

- Both models are globally equivalent and in agreement about the existence of an achromatic $\pm\pi/2$ phase difference between the transmitted and reflected beams. We note however that quantum physics essentially tends to consider the BS as a black-box system fulfilling certain external constraints, while the classical model provides better understanding of the involved physical mechanisms and of their limitations. Alternatively, one may also talk about a "top-down" approach in opposition to a "bottom-up" description.

- Finally, it is striking to note that classical physics can only explain the achromatic $\pi/2$ phase-shift by multiple interference effects (no classical particle model could succeed in it). There may be seen a clear manifestation of the wave nature of light, that must coexist with its corpuscle-like aspect when beamsplitters are precisely employed in experiments demonstrating the existence of the photon. In other words wave-like properties would be inescapable to justify the particle nature of light. Solving this apparent paradox requires reexamining some of these well-known experiments, which is precisely the scope of the next section.

## 3 EXPERIMENTAL SETUPS

In this section are reviewed three emblematic quantum optics experiments mostly conceived for demonstrating the existence of the photon, namely the historical Hanbury Brown and Twiss experiment used in coincidence counting mode (§ 3.1) and the Mach-Zehnder interferometer (§ 3.2). For each of them the experimental results are discussed and possible reinterpretations inspired from classical wave theory are presented. The Hong-Ou-Mandel experiment [2] will be discussed in future publication.

### 3.1 Coincidence counting behind a beamsplitter

#### 3.1.1 The Hanbury Brown and Twiss (HBT) experiment

Historically, the HBT experiment was probably the first aiming at measuring photon correlations [15]. Originally intended to validating the principle of intensity interferometry [16], it gave birth to the theory of fourth-order quantum interference [17] and to a vast family of modern variants operating in coincidence counting mode[1]. A very general and schematic view of such type of experimental setup is depicted in Figure 5: the beamsplitter (BS) is illuminated by a highly coherent light source (e.g. laser, atomic cascade, spontaneous parametric down-conversion from nonlinear crystal). The light amplitude $A_1$ impinging the BS is divided into one transmitted beam and one reflected beam. Two detectors D1 and D2 transform the light intensities into electrical signals $I_{T1}$ and $I_{R1}$. A third signal $C_{RT}$ is produced by electrically correlating $I_{T1}$ and $I_{R1}$ over a short integration time $\tau$ (typically a few nanoseconds). The normalized correlation factor $C$ is usually defined as:

$$C = \frac{C_{RT}}{I_{T1} + I_{R1}}. \qquad (10)$$

One of the most remarkable results obtained from such experiments was reported by GRA [1]. They revealed a significant anti-correlation at low flux level (i.e. $C \to 0$ when $I_{T1} + I_{R1} \to 0$) that was interpreted as confirming the corpuscle-like behavior of the photon[2] (the canonic photon is indivisible and can go one way only – either transmitted or reflected, but never both at the same time as sketched in the right side of Figure 5). But an alternative interpretation relaying on the BS optical properties can be envisioned, as explained in the next sib-section.

---

[1] Photon coincidence counters can actually be considered as photon correlators having very short integration times.
[2] It should be noted that no marked anti-correlation was observed on the historical HBT experiment [15]. Little is know, however about, the optical characteristics of the employed BS (a half-silvered mirror).

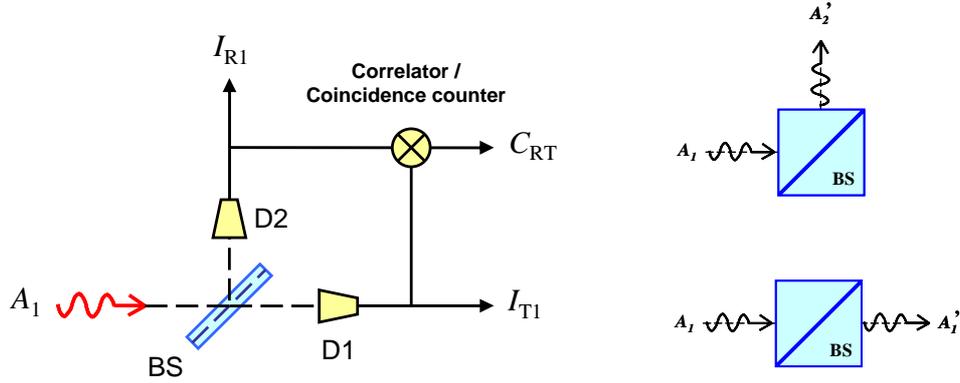

Figure 5: Left: schematic representation of HBT experiment. Right: the undivided photon can only be reflected or transmitted by the BS.

### 3.1.2 Classical correlation model

Here below is described an elementary correlation model compatible with the GRA experimental results, only making use of classical wave optics and of the achromatic phase difference introduced by the BS. In this context a monochromatic electromagnetic field of wavenumber $k = 2\pi/\lambda$ impinging on the beamsplitter can be written as:

$$A_1(t,k) = A_0 \exp\left[-ikc\left(t - \frac{z}{c}\right)\right], \tag{11}$$

where $t$ designates time, $c$ is the speed of light, $z$ the spatial coordinate along the propagation axis, and $A_0$ is an arbitrary constant. We assume the BS in Figure 5 to be lossless and having equal transmitted and reflected intensities, i.e. satisfying to Eq. 9. Using O' as OPD reference point (see Figure 3), the transmitted and reflected amplitudes can be expressed as:

$$A_{T1}(t,k) = \frac{A_0}{\sqrt{2}} \exp\left[-ikc\left(t - \frac{z_{D1}}{c}\right)\right] \quad \text{and:} \quad A_{R1}(t,k) = i\frac{A_0}{\sqrt{2}} \exp\left[-ikc\left(t - \frac{z_{D2}}{c}\right)\right], \tag{12}$$

where the $i$ factor stands for the intrinsic $\pi/2$ phase-shift. Practically the photons are emitted in series of very short time pulses, to which can be associated an equivalent coherence time $\tau_c$ and coherence length $L_c = c\,\tau_c$. From basic wave optics theory this corresponds to a spectral enlargement of the incident radiation $\delta\lambda = \lambda^2/L_c$, or alternatively $\delta k = 2\pi/L_c = 2\pi/c\tau_c$. The instantaneous correlation signal $C_{RT}^{\delta k}(t)$ can then be estimated as the integral of the product $A_{T1}(t,k)A_{E1}(t,k)$ over the wavenumber range $[k - \delta k/2, k + \delta k/2]$:

$$C_{RT}^{\delta k}(t) = \frac{1}{2\delta k}\int_{k-\delta k/2}^{k+\delta k/2} A_{T1}(t,k')A_{R1}(t,k')dk' = i\frac{A_0^2}{2\delta k}\int_{k-\delta k/2}^{k+\delta k/2} \exp\left[-2ik'c\left(t - \frac{z_{D1} + z_{D2}}{c}\right)\right]dk'. \tag{13}$$

It must be noticed that the integral is summed over complex amplitudes rather than to intensities, as was assumed already in Ref. [4] for proposing a classical demonstration of Bell's inequalities violation. Now changing variable $k'$ to $k'' = k' - k$ allows developing Eq. 13 into[1]:

---

[1] This expression could also be derived by use of Fourier transformation.

$$C_{RT}^{\delta k}(t) = i\frac{A_0^2}{2\delta k}\exp\left[-2ikc\left(t - \frac{z_{D1}+z_{D2}}{c}\right)\right]\int_{-\delta k/2}^{+\delta k/2}\exp\left[-2ik''c\left(t - \frac{z_{D1}+z_{D2}}{c}\right)\right]dk''$$

$$= i\frac{A_0^2}{2}\exp\left[-2ikc\left(t - \frac{z_{D1}+z_{D2}}{c}\right)\right]\sin c\left[\delta kc\left(t - \frac{z_{D1}+z_{D2}}{c}\right)\right], \quad (14)$$

where sinc denotes the sine cardinal function $\sin c(u) = \sin u / u$. Assuming that both detectors D1 and D2 are set symmetric with respect to the BS plane and that the time integration window has a half-width $\tau$ centered on $t = (z_{D1}+z_{D2})/c$, the recorded correlation signal $C_{RT}$ may be expressed as:

$$C_{RT} = \frac{1}{2\tau}\int_{-\tau+(z_{D1}+z_{D2})/c}^{+\tau+(z_{D1}+z_{D2})/c}\left[\operatorname{Re}al\left(\frac{C_{RT}^{\delta k}(t)}{A_0^2/2}\right)\right]^2 dt = \frac{1}{2\tau}\int_{-\tau}^{+\tau}\sin^2(2kct)\sin c^2(\delta kct)dt. \quad (15)$$

The last assumption consists in approximating the sinc function by a "step" function equal to unity if $-\tau_c \le \tau \le +\tau_c$ and to zero elsewhere, i.e. in matching the step function width to the first lobe of the sinc function. Moreover, in such kind of coincidence-count measurements the integration time $\tau$ is typically adjusted to twice the coherence time, i.e. $\tau = 2\tau_c$. Hence after elementary trigonometric manipulations:

$$C_{RT} \approx \frac{1}{2\tau}\int_{-\tau_c}^{+\tau_c}\sin^2(2kct)dt = \frac{1-\sin c(2kc\tau)}{4}. \quad (16)$$

Using similar mathematical developments for the output intensities, one also finds that $I_{T1} = |A_{T1}|^2 = [1+\sin c(kc\tau)]/2$ and $I_{R1} = |A_{R1}|^2 = [1-\sin c(kc\tau)]/2$. Thus from Eq. 10 the normalized correlation factor $C$ is finally equal to:

$$C = [1-\sin c(2kc\tau)]/4. \quad (17)$$

Since $\sin c(0) = 1$, it follows that the correlation factor tends towards zero when decreasing the integration time $\tau$. This is clearly due to the presence of the sinc function, itself being a consequence of the achromatic $\pi/2$ phase difference introduced by BS. This result is in contradiction with GRA, who state that $C$ should stay constant in the frame of classical physics, and that only quantum physics could explain the BS anti-correlation effect. However Eq. 17 has actually been derived from pure classical wave optics and involves no photon statistics.

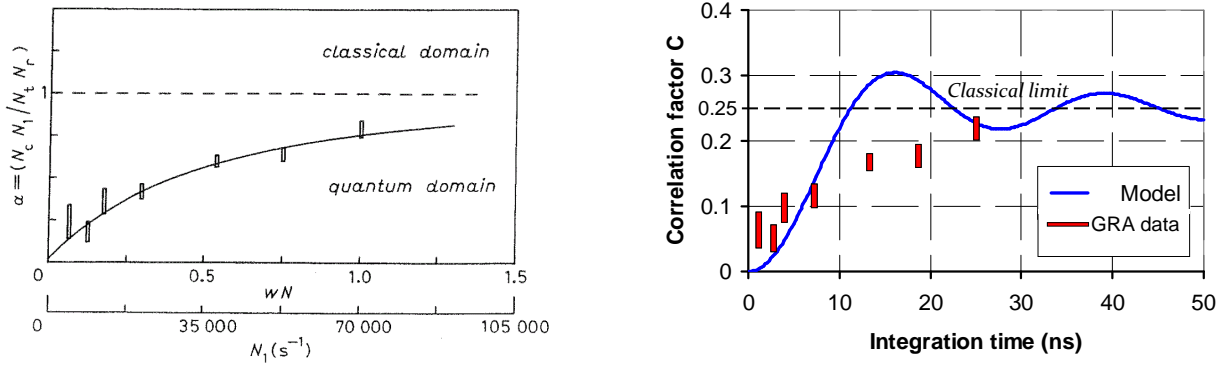

Figure 6: Plot of the correlation factor C as function of integration time $\tau$ (right panel), compared with GRA original experimental results (left panel, reproduced from Ref. [1]).

In Figure 6 is shown a plot of the correlation factor $C(\tau)$ as a function of integration time $\tau$, to which the original results from Ref. [1] were fitted approximately. Although they are not in perfect agreement, the general aspect of the curves suggests that a non-quantum explanation of photon anticcorrelations is possible. Moreover, the main discrepancy between the simplified model and experimental data occurs at the first negative lobe of the sin*c* function (from 11 to 22 ns), and the presence of this function in Eq. 17 essentially results from crude approximations aiming at defining an analytical expression of the correlation factor. Hence it is likely that more realistic results will be obtained through numerical modelling.

### 3.2 The Mach-Zehnder interferometer

*3.2.1 Wave optics description*

The Mach-Zehnder (MZ) is a well-known type of interferometer originally conceived for comparing two parallel and collimated optical beams by means of their phase difference. The optical setup is described schematically in Figure 7. It is made of two beamsplitters BS1 and BS2, and of two flat mirrors M1 and M2 redirecting the light beams splitted by BS1 to the mixing or "recombining" BS2. From it are emerging two complex amplitudes $A'_1$ and $A'_2$, respectively named constructive and destructive outputs. As illustrated in Figure 7, one may distinguish two different basic configurations of the MZ interferometer:

- The "non symmetric" configuration (left part of figure), which is probably the most common. Here the BS are two identical half-silvered mirrors, being arranged so that the four possible light paths from the input to the output ports of the interferometer experience the same number of grass crossing. Only a single reflection onto the semi-reflective coatings is contributing to interference formation, and multiple interferences inside the glass plates are undesirable since they would produce spurious interference patterns. Practically, they can be eliminated either by making use of slightly wedged glass plates (as depicted on the figure) or of high performance anti-reflection coatings. In this configuration the Fresnel reflection coefficients at the semi-reflective layers of BS1 and BS2 are of opposite signs, being either internal or external to the glass and thus π-phase-shifted one with respect to the other [18-19]. Obviously the beamsplitter model of section 2.2 is not directly applicable to this MZ configuration, nor the quantum matrix representation of Eq. 3a (one may use Eq. 3b instead).

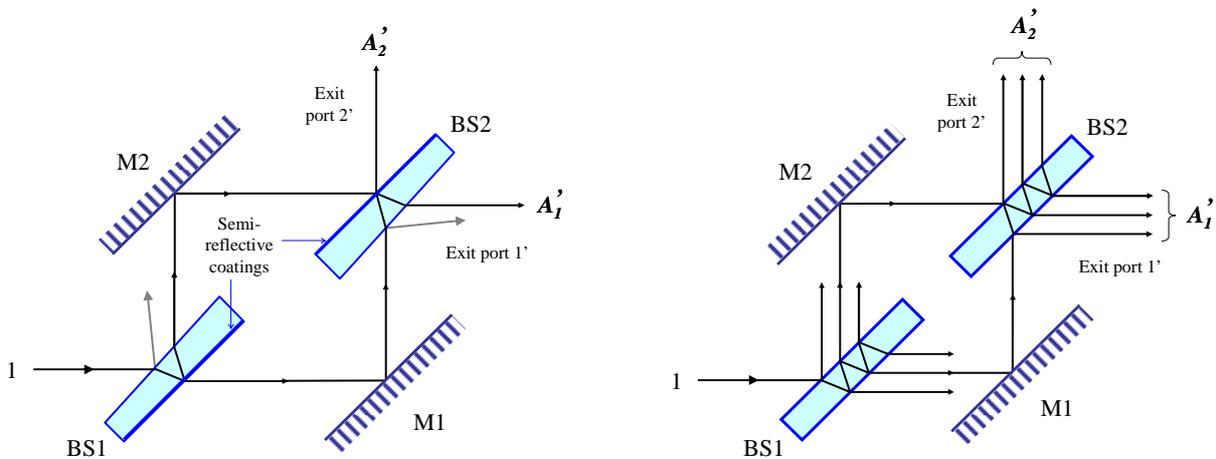

Figure 7: Two basic configurations of the MZ interferometer: non symmetric (left panel) and symmetric (right panel).

- The "symmetric" configuration (right part of figure), where BS1 and BS2 are two identical and symmetric plane and parallel plates as described in section 2.2 (the cube beamsplitter depicted in Figure 2 stands for a good example). Hence the FP multi-interference effect participates to the output complex amplitudes, Eqs. 4-9 are fully applicable, and there exist an achromatic quarter wave phase difference between $A'_1$ and $A'_2$. Employing the same notations as in sub-section 2.2, the expressions of the amplitudes and intensities at the output ports of the MZ interferometer are:

|  | Complex amplitude | Intensity |
|---|---|---|
| Output port 1' (constructive) | $A'_1 = 2 A_{T1} A_{R1}$ | $I'_1 = |A'_1|^2 = 4 |A_{T1}|^2 |A_{R1}|^2$ |
| Output port 2' (destructive) | $A'_2 = A_{T1} A_{T1} + A_{R1} A_{R1}$ | $I'_2 = |A'_2|^2 = |A_{T1} A_{T1} + A_{R1} A_{R1}|^2$ |

From Eqs 4a and 6, it is not very difficult to demonstrate that:
$$A_{T1} A_{R1}^* + A_{R1} A_{T1}^* = 0 \,^1. \tag{18}$$

Therefore the total energy transmitted by the system is conserved and equal to $I'_1 + I'_2 = 1 + (A_{T1} A_{R1}^* + A_{R1} A_{T1}^*)^2 = 1$. Moreover, the phase difference between both output arms of the interferometer are evaluated as the argument of $A'_2 A'^*_1$ and found to be:

$$\phi_2 - \phi_1 = Arg[A'_2 A'^*_1] = Arg[2(I_{T1} - I_{R1}) A_{T1} A_{R1}^*] = Arg[A_{T1} A_{R1}^*] = \pm \frac{\pi}{2}, \tag{19}$$

that is quite similar to those of the single beamsplitter in Eq. 7[2]. It follows that in the frame of classical wave optics the symmetric MZ interferometer may be interpreted as a synthetic beamsplitter having similar properties of energy conservation and achromatic phase-shift.

As previously, numerical simulations of the MZ output beans showing their intensities and phases were carried out, whose results are presented in Figure 8. In panel A are shown the intensity variations $I'_1$ and $I'_2$ as function of the BS internal phase $\varphi$ (common to both BS1 and BS2). The output phase curves $\phi_1(\varphi)$, $\phi_2(\varphi)$ and their difference $\phi_2 - \phi_1$ are displayed in panel B, where the latter is found compliant with the prediction of Eq. 18. These results confirm that a static MZ interferometer globally behaves as one single BS. We note finally that the conditions for achieving equal BS output intensities (i.e. $\varphi \approx 0.115\pi$ or $0.885\pi$ [$\pi$] as in § 2.2) now corresponds to the destructive MZ output ($I'_2 = 0$ in panel A), and to phase jumps of $\phi_2(\varphi)$ in panel B.

*3.2.2 Optical path difference (OPD) modulation*

Practically the Mach-Zehnder interferometer is often used with internal OPD modulation, for example operating as a Fourier transform spectrometer. The basic principle consists in adding a variable optical path $\delta$ between both interferometer arms, for example by means of glass plates of variable optical thickness as sketched in Figure 9 or of optical delay lines. Being equivalent to a time delay $\delta/c$, this OPD can be translated into a phase-shift $\varphi = \delta k = 2\pi\delta/\lambda$. Unlike the BS internal phase difference, the latter is however not achromatic. Assuming that the OPD modulation is introduced onto BS1 transmitted beam (see Figure 9), the complex amplitudes at the output ports of the interferometer become:

Output port 1' $\quad A'_1 = A_{T1} A_{R1} [1 + \exp(ik\delta)]$
Output port 2' $\quad A'_2 = A_{T1} A_{T1} \exp(ik\delta) + A_{R1} A_{R1}$

Here again and using relation 18, it is not very difficult to verify that the principle of energy conservation is respected, since $I'_1 + I'_2 = 1 + A_{T1} A_{R1}^* [A_{R1} A_{T1}^* + A_{T1} A_{R1}^*] \exp(ik\delta) + A_{R1} A_{T1}^* [A_{R1} A_{T1}^* + A_{T1} A_{R1}^*] \exp(-ik\delta) = 1$. It follows that:

$$I'_1 = 4 |A_{T1} A_{R1}|^2 \cos^2(k\delta/2) \quad \text{and} \quad I'_2 = 1 - 4|A_{T1} A_{R1}|^2 + 4|A_{T1} A_{R1}|^2 \sin^2(k\delta/2) \tag{20}$$

Also, the phase difference between exit ports 1' and 2' is developed as:

$$\phi_2 - \phi_1 = Arg\left[ 2\cos(k\delta/2) A_{T1}^* A_{R1} \left\{ |A_{R1}|^2 \exp(-ik\delta/2) - |A_{T1}|^2 \exp(ik\delta/2) \right\} \right],$$

---

[1] For lossless beamsplitters this relation is equivalent to Eq. 2b derived from the black-box formalism of quantum optics.
[2] We note that $\phi_2 - \phi_1$ is undetermined when $I_{T1}$ and $I_{R1}$ are equal. This is consistent with the fact that $I'_2 = 0$ in that case.

which for intensity-balanced beamsplitters ($|A_{T1}|^2 = |A_{R1}|^2 = 0.5$) reduces to:

$$\phi_2 - \phi_1 = Arg\left[-2i|A_R|^2 \sin(k\delta) A_{T1}^* A_{R1}\right] = 0 \quad [\pi]. \tag{21}$$

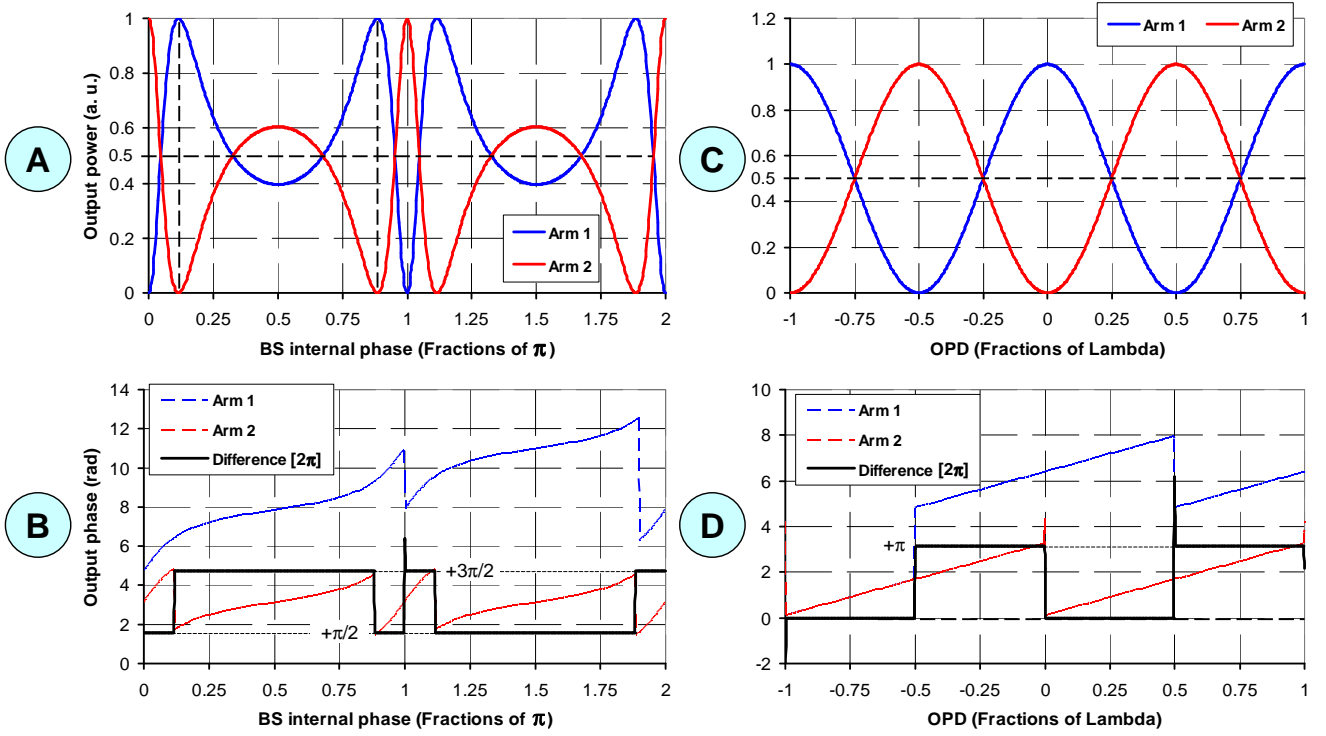

Figure 8: Intensity and phases curves of the MZ output beams shown as function of BS internal phase $\varphi$ in panels A and B, and of OPD modulation $\delta$ in panels C and D.

Also, the phase difference between exit ports 1' and 2' is developed as:

$$\phi_2 - \phi_1 = Arg\left[2\cos(k\delta/2) A_{T1}^* A_{R1} \left\{|A_{R1}|^2 \exp(-ik\delta/2) - |A_{T1}|^2 \exp(ik\delta/2)\right\}\right],$$

which for intensity-balanced beamsplitters ($|A_{T1}|^2 = |A_{R1}|^2 = 0.5$) reduces to:

$$\phi_2 - \phi_1 = Arg\left[-2i|A_R|^2 \sin(k\delta) A_{T1}^* A_{R1}\right] = 0 \quad [\pi]. \tag{21}$$

These formulae are illustrated in Figure 8, showing the output intensity curves $I'_1(\delta)$ and $I'_2(\delta)$ as function of the OPD modulation (panel C) and the relative phase-shift between them (panel D) for the case of a lossless and intensity-balanced beamsplitter. This last result may look counter-intuitive because $\phi_2 - \phi_1$ is not equal to $\pi$ constantly (phase jumps occur either when $I'_1$ or $I'_2 = 0$). But in reality the misleading idea that the exit beams of a MZ interferometer are always in phase opposition probably originates from the non symmetric configuration of Figure 7, where $\phi_2 - \phi_1$ can be mistaken for the $\pi$-phase-shift of Fresnel's reflection coefficients at BS2 (internal and external to the glass).

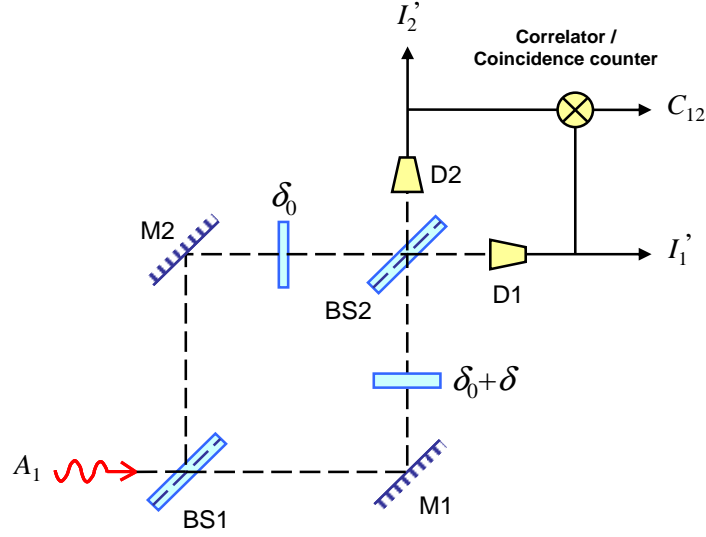

Figure 9: Schematic diagram of the Mach-Zehnder interferometer in coincidence counting mode.

### 3.2.3 Quantum physics description

The quantum physics description of the Mach-Zehnder interferometer equipped with lossless and symmetric beamsplitters does not fundamentally departs from the classical approach. It makes use of the same type of matrices as in § 2.1, arranged in the following order [8][1]:

$$\begin{bmatrix} A'_2 \\ A'_1 \end{bmatrix} = M_{MZ} \begin{bmatrix} A_2 \\ A_1 \end{bmatrix} = M_{BS} M_\delta M_{BS} \begin{bmatrix} A_2 \\ A_1 \end{bmatrix} = \frac{1}{2} \begin{bmatrix} 1 & i \\ i & 1 \end{bmatrix} \begin{bmatrix} \exp(ik\delta) & 0 \\ 0 & 1 \end{bmatrix} \begin{bmatrix} 1 & i \\ i & 1 \end{bmatrix} \begin{bmatrix} A_2 \\ A_1 \end{bmatrix}, \qquad (22)$$

where $M_{BS}$ is the BS matrix from Eq. 3a and $M_\delta$ stands for OPD modulation inside the interferometer (here the original matrix relation in Ref. [8] has been slightly modified, due to different notation conventions from the present paper). Then for an input state vector $[A_1, A_2] = [1,0]$, Eq. 22 readily leads to $I'_1 = \cos^2(k\delta/2)$, $I'_2 = \sin^2(k\delta/2)$ and $\phi_2 - \phi_1 = 0\ [\pi]$, being in full compliance with the predictions of the classical model (Eqs. 20-21 with $|A_{T1}|^2 = |A_{R1}|^2 = 0.5$).

### 3.2.4 Quantum physics experiments

Though essentially employed as a metrology tool in optics or gas dynamics, the Mach-Zehnder interferometer has regularly been involved in quantum optics experiments questioning the nature of light. These experiments may be divided into two categories:

- Single photon interference, meaning that only one side of BS1 is illuminated by a source of photons. Fourth-order interference is realized by coincidence counting measurement at the exit ports of the interferometer, as was the single BS in the first step of the GRA experiment (see § 3.1.1). Typical or related examples of quantum MZ experiments include testing of Heisenberg's uncertainty principle [21-22], the Franson interferometer investigating on hidden-variable theory [23] and the second step of the GRA experiment [1].

- Two-photon interference, where both faces of BS1 are simultaneously illuminated. These MZ experiments essentially aim at discriminating second and fourth-order interference as function of the employed light source [20][24].

---

[1] A similar type of relationship can be established for the non symmetric MZ configuration, see Ref. [20], Appendix B.

Herein is only discussed single photon interference, and particularly the second step of the GRA experiment. After measuring photons anti-correlation (see § 3.1.1), the authors built a MZ interferometer around the same beamsplitter BS1 as used previously, and observed at same light levels the characteristic intensity modulations $I'_1(\delta)$ and $I'_2(\delta)$ of classical wave optics. They interpreted this result as a plain demonstration of the complementarily principle, but in light of the alternative correlation model presented in sub-section 3.1.2, one may alternatively consider their results to be in favor of the "wave-only" nature of light, even at low light level and minimal photons number. More generally, it should also be noted that, though presented as a decisive proof of the corpuscular nature of light, such experimental results can only be explained by adding wave-like properties (i.e. phase functions interpreted as probability amplitudes) to the quantum formalism described in most of the reference papers.

## 4 CONCLUSION

In the first half of this communication were reexamined the quantum and classical wave optics theories of the beamsplitter. It was shown that both models are globally in agreement and account for the existence of an achromatic ±π/2 phase difference between the transmitted and reflected electric fields. Despite their apparent equivalence, the wave optics model looks much more convenient for introducing typical BS characteristics and defects. More generally, it was noted that quantum physics essentially considers the BS as a black-box system, while the classical model provides deeper understanding of the physical mechanism generating the achromatic phase-shift, that is a classical multiple interference effect.  properties are necessary for full BS description"" or photons

Such notionhe second part of this study. Here were reviewed a few impressive experiments demonstrating the existence of the photon and the corpuscular nature of light. Those included the HBT coincidence counting test and the Mach-Zehnder interferometer. For each of them was developed a minimal wave optics model based on very simple hypotheses, tending to reproduce their experimental results without requiring the intervention of light particles.  of fourth-order interference phenomena strictly belonging to the sole realm of quantum physics. It must be emphasized, however, that only a very small fraction of today's realized quantum experiments on optical correlations and interference was reviewed in this paper. Other interferometric tests will be the scope of a future paper [25], including the Hong-Ou-Mandel experiment and refinements of the herein presented wave optics model.

## REFERENCES

[1] P Grangier, G Roger, A Aspect, "Experimental evidence for a photon anticorrelation effect on a beam splitter: a new light on single-photon interferences," Europhysics Letters 1, p. 173-179 (1986).
[2] C. K. Hong; Z. Y Ou,. L. Mandel, "Measurement of subpicosecond time intervals between two photons by interference," Phys. Rev. Lett. vol. 59, p.: 2044-2046 (1987).
[3] A. Aspect, J. Dalibard, G. Roger, "Experimental test of Bell's inequalities using variable analyzers," Physical Review Letters vol. 49, p. 1804-1807 (1982).
[4] F. Hénault, "Can violations of Bell's inequalities be considered as the final proof of quantum physics ?," Proceedings of the SPIE vol. 8832, n° 88321J (2013).
[5] V. Degiorgio, "Phase shift between the transmitted and the reflected optical fields of a semireflecting lossless mirror is π/2," American Journal of Physics vol. 48, p. 81-82 (1980).
[6] Z. Y. Ou, L. Mandel, "Derivation of reciprocity relations for a beam splitter from energy balance," American Journal of Physics vol. 57, p. 66-67 (1989).
[7]  A. Zeilinger, "General properties of lossless beam splitters in interferometry, American Journal of Physics vol. 49, p. 882-883 (1981).
[8] C. H. Holbrow, E. Galvez, M. E. Parks, "Photon quantum mechanics and beam splitters," American Journal of Physics vol. 70, p. 260-265 (2002).

[9] H. Fearn, R. Loudon, "Quantum theory of the lossless beam splitter," Optics Communications vol. 64, p. 485-490 (1987).
[10] R. A. Campos, B. E. Saleh, M. C. Teich, "Quantum-mechanical lossless beam splitter: SU(2) symmetry and photon statistics," Physical Review A vol. 40, p. 1371-1384 (1989).
[11] J. H. Shapiro, K.-X. Sun, "Semiclassical versus quantum behavior in fourth-order interference," JOSA B vol. 11, p. 1130-1141 (1994).
[12] P. Yeh, *Optical Waves in Layered Media*, chapter 4 (John Wiley and Sons 1998).
[13] M. Born, E. Wolf, *Principles of Optics*, 6$^{th}$ Edition, section 7.6 (Cambridge University 1999).
[14] M. W. Hamilton, "Phase shifts in multilayer dielectric beam splitters," American Journal of Physics vol. 68, p. 186-191 (2000).
[15] R. Hanbury Brown, R. Q. Twiss, "Correlations between photons in two coherent beams of light," Nature vol. 177, p. 27-29 (1956).
[16] R. Hanbury Brown, R. Q. Twiss, "A test of a new type of stellar interferometer on Sirius," Nature vol. 178, p. 1046-1048 (1956).
[17] J. G. Rarity and P. R. Tapster, "Fourth-order interference in parametric down conversion," JOSA B vol. 6, p. 1221-1226 (1989).
[18] K. P. Zetie, S. F. Adams, R. M. Tocknell, "How does a Mach-Zehnder interferometer work?," Physics Education vol. 35, p. 46-48 (2000).
[19] A. Carlotti, "Utilisation d'un interféromètre de Mach-Zehnder pour la mise en oeuvre de coronographes achromatiques," PhD Thesis, Université Nice-Sophia Antipolis (2009).
[20] R. A. Campos, B. E. Saleh, M. C. Teich, "Fourth-order interference of joint single-photon wave packets in lossless optical systems," Physical Review A vol. 42, p. 4127-4137 (1990).
[21] S. Dürr, G. Rempe, "Can wave-particle duality be based on the uncertainty relation?," Am. J. Phys. 68, p. 1021-1024 (2000).
[22] H.-Y. Liu, J.-H. Huang, J.-R. Gao, M. S. Zubairy, S.-Y. Zhu, "Relation between wave-particle duality and quantum uncertainty," Phys. Rev. A vol. 85, n° 022106 (2012).
[23] J. D. Franson, "Bell inequality for position and time," Phys. Rev. Lett. vol. 62, p. 2205-2208 (1989).
[24] Z. Y. Ou, X. Y. Zou, L. J. Wang, L. Mandel, "Experiment on nonclassical fourth-order interference," Phys. Rev. A vol. 42, p. 2957-2965 (1990).
[25] The author is preparing a paper entitled "Quantum physics and the beam splitter mystery II. Interference experiments."